\documentclass[prx,twocolumn,preprintnumbers,amsmath,amssymb]{revtex4}

\usepackage{graphicx}% Include figure files
\usepackage{bm}% bold math
\usepackage{color}% bold math

\begin{document}

%\preprint{APS/123-QED}

\title{Lattice dynamics for isochorically heated metals: A model study}
\author{Shota Ono}
\email{shota\_o@gifu-u.ac.jp}
\affiliation{Department of Electrical, Electronic and Computer Engineering, Gifu University, Gifu 501-1193, Japan}

\begin{abstract}
The electron-excitation induced bond strength variations in metals have been predicted from density-functional theory calculations and observed experimentally, while the microscopic mechanism has yet to be elucidated. Here, we present a minimal model that reproduces the phonon hardening and softening for fcc- and bcc-structured metals as a result of the electron thermal excitation. We explain why the phonon mode softens at the N point for bcc-structured metals.
\end{abstract}

%\pacs{61.48.-c, 72.80.Rj, 73.20.Fz}
%61.48.-c: Structure of fullerenes and related hollow and planar molecular structures
%72.80.Rj: Fullerenes and related materials / Conductivity of specific materials
%73.20.Fz: Weak or Anderson localization / Electron states at surfaces and interfaces

\maketitle

%%%%%%%%%%%%%%%%%%%%%%%%%%%%%%%%%%%
\section{Introduction}
%%%%%%%%%%%%%%%%%%%%%%%%%%%%%%%%%%%
The effect of strong nonequilibrium condition between electrons and phonons on solid state properties has been investigated both experimentally and theoretically. It influences not only thermodynamic properties, such as the electron specific heat and density-of-states \cite{lin}, but also the lattice dynamics spectra (i.e., phonon dispersion relations). Since the values of the force constants for ions are determined by the adiabatic potential that is a sum of the ion-ion repulsive and the ion-electron-ion interaction potential energies, it is possible to manipulate the lattice dynamics spectra by tuning the electron-mediated interaction potential. Based on density-functional theory (DFT) calculations, Recoules {\it et al}. have predicted that the phonon frequencies of Au increase over the entire Brillouin zone (BZ) when the electron temperature $T_{\rm e}$ is increased up to several eV \cite{recoules}. This is understood as a decrease in the electron-ion screening as a result of the thermal excitation of $5d$-electrons located below the Fermi level by a few eV. The femtosecond pump-probe technique has confirmed the bond hardening as an increase in the melting temperature of Au \cite{ernstorfer}, which paves the way for understanding the fundamental properties of warm-dense aluminum \cite{leguay}, copper \cite{cho}, molydbenum \cite{dorchies}, and electron gas \cite{groth}. 

By this argument recent studies based on DFT calculations in Ref.~\cite{minakov,yan,harbour,bottin} is interesting; Even if $d$-electrons are absent in a system, a noticeable change in the phonon dispersion relations has been predicted when $T_{\rm e}$ increases. For example, fcc-structured metals such as Al show a phonon hardening over the entire BZ, while bcc-structured metals such as Na show a phonon softening at the point N within the BZ \cite{yan}. A similar conclusion has been reached in a recent study \cite{harbour}, where the neutral pseudoatom model developed from DFT and molecular dynamics simulations has been used. Interestingly, Bottin {\it et al}. have shown that for both Al and Au crystals the monovacancy formation enthalpy increases with $T_{\rm e}$ (i.e., the bond hardening), while its origin is different: The largest contribution to the stress is from the kinetic energy part for Al and the pseudopotential (ion-electron potential) energy part for Au \cite{bottin}. Since Al and Na are a typical free-electron metal in the ground state, as well as the supporting evidence for high $T_{\rm e}$ \cite{bottin}, it must be possible to develop a simple model to understand such a crystal structure dependence of the phonon property. 

%Warm dense matter is an extreme state of solids: the electron temperature is comparable to the Fermi temperature, while the lattice temperature is left to the room temperature. The phonon hardening of rare metals such as warm dense gold have been reported. This is understood as the decreased electron-ion screening as a result of the excitation of the $5d$-electrons located below the Fermi level by a few eV . By this argument recent theoretical work in Ref.~\cite{yan} is interesting: Even if $d$-electron is absent, noticeable change in the phonon dispersion relation as a function of the electronic temperature has been predicted. For example, fcc-structured metals such as Al show a phonon hardening, while bcc-structured metals such as Na show a phonon softening. Such a crystal structure dependence of the phonon property would be understood by more simple models. 

In this paper, we construct a minimal model for phonons in isochorically heated nearly free-electron metals and calculate the $T_{\rm e}$-dependence of the phonon dispersion relations. The phonon hardening and softening occur in fcc- and bcc-structured metals, respectively, which are consistent with DFT based calculations \cite{yan,harbour}. The phonon hardening originates from a significant increase in the force constant for the first nearest neighbor (NN) sites, while the phonon softening at the N point in bcc-structured metals originates from a delicate balance between force constants for the first and second NN sites.

%%%%%%%%%%%%%%%%%%%%%%%%%%%%%%%%%%%
\section{Formulation}
%%%%%%%%%%%%%%%%%%%%%%%%%%%%%%%%%%%
To compute the phonon dispersion relations in metals, we extend the theory of the lattice dynamics for simple metals at $T_{\rm e}=$0 K \cite{hartmann} to the case at $T_{\rm e}\ne$0 K. We consider a simple metal that consists of ions and conducting electrons. Each ion and electron have charges $Ze$ and $-e$, respectively, where $Z$ is the valence of the ion. With a charge neutrality condition, the number of electrons is uniquely determined when that of ions is given. We assume that the total potential energy between ions separated by a distance $R$ is 
\begin{eqnarray}
V_{\rm tot}(R)
 &=& v_{\rm d}(R)
 + v_{\rm ind}(R),
\label{eq:ion-ion}
\end{eqnarray}
where $v_{\rm d}$ is the direct interaction potential between ions and given by
\begin{eqnarray}
v_{\rm d}(R)
 &=& \frac{Z^2 e^2}{4\pi \varepsilon_0 R}
\label{eq:d}
\end{eqnarray}
with the dielectric constant of vacuum $\varepsilon_0$. $v_{\rm ind}$ in Eq.~(\ref{eq:ion-ion}) is the indirect interaction potential that is derived from the electron-mediated ion-ion interaction. This is written as (see Appendix \ref{app} for the derivation)
\begin{eqnarray}
v_{\rm ind}(R)
 &=& 
 \int_{0}^{\infty} dq C(q) \frac{\sin (qR)}{qR}
\label{eq:ind}
\end{eqnarray}
with the wavenumber $q$. The kernel $C(q)$ in Eq.~(\ref{eq:ind}) is 
\begin{eqnarray}
C(q)
 &=& - \left(\frac{\varepsilon_0 q^4}{2\pi^2 e^2}\right) v_{\rm ps}^{2}(q)
 \frac{\chi(q,T_{\rm e})}{1+[1-G(q)]\chi(q,T_{\rm e})},
\label{eq:ind2}
\end{eqnarray}
where $v_{\rm ps}(q)$ is the Fourier component of the model pseudopotential. $\chi(q,T_{\rm e})$ is the $T_{\rm e}$-dependent response function and explicitly written as
\begin{eqnarray}
\chi(q, T_{\rm e})
 &=& \frac{4}{\pi k_{\rm F}a_{\rm B}y^2} 
 \int_{0}^{\infty} dx \frac{x}{y} f(x,T_{\rm e}) 
 \ln \left\vert \frac{2+y/x}{2-y/x} \right\vert
 \nonumber\\
 \label{eq:response}
\end{eqnarray}
with the Fermi wavenumber $k_{\rm F}$, the Bohr radius $a_{\rm B}$, $x=k/k_{\rm F}$, $y=q/k_{\rm F}$, and the Fermi-Dirac distribution function
\begin{eqnarray}
 f(x,T_{\rm e}) = \left[ e^{(\varepsilon_{\rm F}x^2 - \mu)/(k_{\rm B}T_{\rm e})}\right]^{-1}
 \label{eq:fermi}
\end{eqnarray}
with the Fermi energy $\varepsilon_{\rm F}$ at $T_{\rm e}=0$ K, the chemical potential $\mu$, and the Boltzmann constant $k_{\rm B}$. When $T_{\rm e}=0$ K, Eq.~(\ref{eq:response}) can be reduced to the Hartree formula \cite{hartmann}
\begin{eqnarray}
\chi(q, 0)
 &=& \frac{4}{\pi k_{\rm F}a_{\rm B}y^2} 
\left( 
\frac{1}{2} + 
\frac{4-y^2}{8y}
 \ln \left\vert \frac{2+y}{2-y} \right\vert
 \right).
 \label{eq:response0}
\end{eqnarray}
In our model, the effect of $T_{\rm e}$ (i.e., electron occupation) on the phonon dispersion relations is entered into $\chi(q,T_{\rm e})$. Finally, $G(q)$ in Eq.~(\ref{eq:ind2}) accounts for the effects of exchange and correlation. The model functions $v_{\rm ps}(q)$ and $G(q)$ with material parameters will be given later. 

%%%%%%%%%%%%%%%%%
\begin{figure*}[ttt]
\center
\includegraphics[scale=0.5]{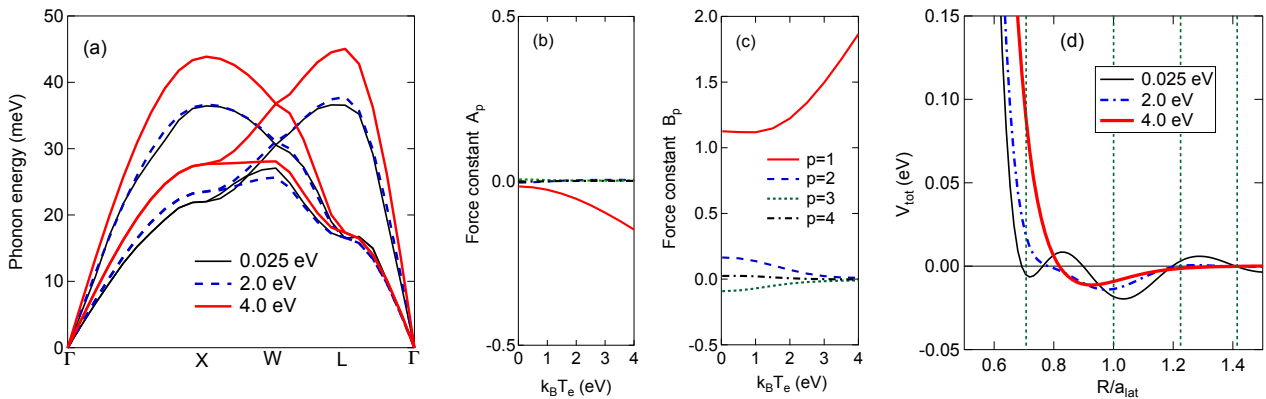}
\caption{\label{fig1}(a) The phonon dispersion relations of fcc-structured Al along symmetry lines for $k_{\rm B}T_{\rm e}=$0.025 (black), 2.0 (blue), and 4.0 (red) eV. $T_{\rm e}$-dependence of the force constants (b) $A_p$ and (c) $B_p$ in units of eV/\AA$^2$. (d) The total potential $V_{\rm tot}$ defined as Eq.~(\ref{eq:ion-ion}). The vertical dotted lines indicate the interatomic distance up to $p=4$. }
\end{figure*}
%%%%%%%%%%%%%%%%%

%%%%%%%%%%%%%%%%%
\begin{figure*}[ttt]
\center
\includegraphics[scale=0.5]{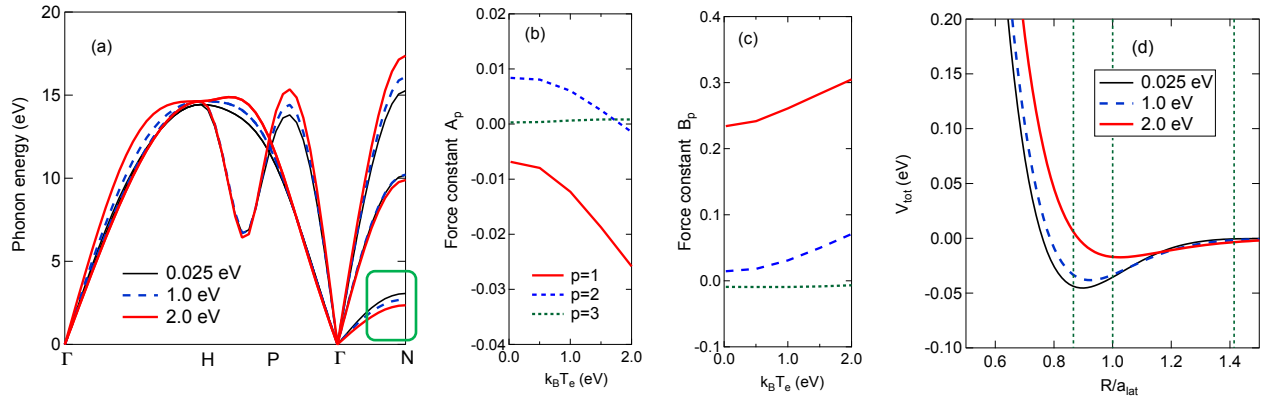}
\caption{\label{fig2} Same as Fig.~\ref{fig1} but for Na. The phonon dispersion relations and $V_{\rm tot}$ are calculated for $k_{\rm B}T_{\rm e}=$0.025, 1.0, and 2.0 eV. The lower frequency phonon at the N point decreases with $T_{\rm e}$ enclosed by a rounded rectangle. The interatomic distance is indicated up to $p=3$.}
\end{figure*}
%%%%%%%%%%%%%%%%%

The phonon dispersion relations for the central potential of Eq.~(\ref{eq:ion-ion}) are calculated by a diagonalization of the dynamical matrix \cite{mermin}
\begin{eqnarray}
 {\cal D} (\bm{q})= \sum_{l} \sin^2 \left( \frac{\bm{q}\cdot \bm{R}_l}{2}\right) 
 \left[ A \bm{1} + B \hat{R}_l \hat{R}_l \right],
 \label{eq:dyn}
\end{eqnarray}
where $\bm{q}$ is the wavevector of phonons, $\bm{R}_l=(R_{lx}, R_{ly}, R_{lz})$ is the $l$th ion position, $\bm{1}$ is the $3\times3$ unit matrix, and $\hat{R}_l \hat{R}_l$ is the dyadic formed from the unit vectors $\hat{R}_l= \bm{R}_l/\vert \bm{R}_l \vert$. $A$ and $B$ are the force constants defined as
\begin{eqnarray}
A &=& \frac{2}{R_l}\frac{dV_{\rm tot} (R)}{dR}\Big\vert_{R=R_l}, 
 \label{eq:A}
\\
B &=& 2\left[ \frac{d^2 V_{\rm tot} (R)}{dR^2}\Big\vert_{R=R_l} 
- \frac{1}{R_l}\frac{dV_{\rm tot} (R)}{dR}\Big\vert_{R=R_l}
\right],
 \label{eq:B}
\end{eqnarray}
where the derivatives of $V_{\rm tot}$ are evaluated at $R_l =\vert \bm{R}_l \vert$. For later use, we define $A_p$ and $B_p$ as the force constant of Eqs.~(\ref{eq:A}) and (\ref{eq:B}) for the $p$th NN ions. The phonon frequencies are given by $\omega = \sqrt{\lambda /M_{\rm ion}}$ with the ion mass $M_{\rm ion}$ and three eigenvalues $\lambda$ of Eq.~(\ref{eq:dyn}). 

%%%%%%%%%%%%%%%%%%%%%%%%%%%%%%%%%%%
\section{Results and discussion}
\label{sec:3}
We study the phonon properties of Al ($Z=3$) and Na ($Z=1$) that show a fcc and bcc structure in the ground state, respectively. The lattice constant is $a_{\rm lat}=4.049$ \AA \ for Al and 4.225 \AA \ for Na. The Wigner-Seitz radius is $r_s=2.07$ for Al and 3.93 for Na (in units of Bohr radius $a_{\rm B}$) \cite{mermin}. The Fermi energy is then calculated to be 11.65 eV for Al and 3.24 eV for Na. For the model potential, we use the Ashcroft pseudopotential
\begin{eqnarray}
 v_{\rm ps} (q) = - \frac{Ze^2}{\varepsilon_0 q^2} \cos(q r_{\rm c}),
 \label{eq:Ashcroft}
\end{eqnarray}
where $r_{\rm c}$ is the cutoff radius, which is set to be 0.5911 \AA \ for Al and 0.8784 \AA \ for Na \cite{ashcroft}. For the correction $G(q)$ for exchange and correlation energies, we use the Hubbard-type function
\begin{eqnarray}
 G(q) = \frac{a q^2}{q^2 + b},
 \label{eq:Hubbard}
\end{eqnarray}
where the parameters of $a$ and $b$ are determined from an analytical formula given in Ref.~\cite{UI}. We have confirmed that the same conclusion (phonon hardening and softening with $T_{\rm e}$) holds when $G(q)$ is set to be zero. We have also performed other bcc-structured crystals (Li, K, Rb, and Cs) and confirmed that the trend of their results is similar to that of Na shown below. 

%%%%%%%%%%%%%%%%%%%%%%%%%%%%%%%%%%%
\subsection{fcc-structured Al}
Figure \ref{fig1}(a) shows the phonon dispersion relations of Al for $k_{\rm B}T_{\rm e}=$0.025, 2.0, and 4.0 eV. A significant increase in the phonon energies is observed at $k_{\rm B}T_{\rm e}=4.0$ eV, which is consistent with the DFT results in Ref.~\cite{yan,harbour}. To understand the phonon hardening driven by an electronic excitation, we show $T_{\rm e}$-dependence of $A_p$ and $B_p$ ($p=1,2,3$, and $4$) in Fig.~\ref{fig1}(b) and \ref{fig1}(c), respectively. The magnitude of $B_1$ starts to increase from $T_{\rm e}\simeq$ 2 eV, while that of $B_p$ for $p\ge 2$ converges to zero, and $A_1$ decreases negatively. These changes are caused by the $T_{\rm e}$-dependence of $V_{\rm tot}(R)$ in Eq.~(\ref{eq:ion-ion}), shown in Fig.~\ref{fig1}(d). When $k_{\rm B}T_{\rm e}=$0.025 eV, $V_{\rm tot}(R)$ shows a Friedel oscillation that originates from the presence of the Fermi surface. As $T_{\rm e}$ increases, the oscillating amplitude becomes weak and thus the value of $V_{\rm tot}$ for $R/a_{\rm lat} >1$ becomes negligibly small, which lead to a significant decrease in $B_p$ for $p\ge 2$. In addition, $V_{\rm tot}$ becomes more repulsive at $p=1$ sites since $v_{\rm d} \gg \vert v_{\rm ind}\vert$. This also leads to an increase in $\vert A_1 \vert$ and $B_1$ that are defined as Eqs.~(\ref{eq:A}) and (\ref{eq:B}), respectively. 

As is clear from Figs.~\ref{fig1}(b) and \ref{fig1}(c), the lattice dynamics up to $k_{\rm B}T_{\rm e}=$ 4.0 eV is almost regulated by $B_1$ only because $B_1 \gg \vert B_p\vert$ ($p=2,3$, and $4$) and $B_1\gg \vert A_p \vert$. A simple analysis, where the contribution from $A_p$ and $B_p$ with $p\ge 2$ is ignored, enables us to understand the phonon hardening phenomena observed above. For example, we focus on the phonon frequency at the X point, at which the phonon frequencies are given by 
\begin{eqnarray}
\omega_{\rm 1,2} = \sqrt{\frac{8A_1+2B_1}{M_{\rm ion}}},
\ \ 
\omega_{\rm 3} = \sqrt{\frac{8A_1+4B_1}{M_{\rm ion}}},
\label{eq:omegaX}
\end{eqnarray}
where $\omega_1$ and $\omega_2$ are the doubly degenerate TA phonon frequencies and $\omega_3$ is the LA phonon frequency. From Eq.~(\ref{eq:omegaX}), it is obvious that the increase in the phonon energy in Fig.~\ref{fig1}(a) is directly related to the increase in $B_1$.

%%%%%%%%%%%%%%%%%%%%%%%%%%%%%%%%%%%
\subsection{bcc-structured Na}
We next investigate the phonon properties of Na. Figures \ref{fig2}(a), \ref{fig2}(b) and \ref{fig2}(c), and \ref{fig2}(d) show the phonon dispersion relations (for $k_{\rm B}T_{\rm e}=$0.025, 1.0, and 2.0 eV), $A_p$ and $B_p$ ($p=1,2$, and $3$), and $V_{\rm tot}(R)$, respectively. As shown in Fig.~\ref{fig2}(a), the phonon energy increases slightly for higher frequency region, while the lowest phonon frequency at the N point decreases with $T_{\rm e}$. Similar softening behavior and an imaginary frequency at the N point have been reported in Ref.~\cite{yan,harbour}. For lower $T_{\rm e}$ the lattice dynamics is almost regulated by $B_1$ again. However, the changes in $A_1$, $A_2$, and $B_2$ in response to $T_{\rm e}$ cannot be negligible. This is because the profile of $V_{\rm tot}$ is different from that in Al (Fig.~\ref{fig1}(d)): The Friedel oscillation is not clearly observed, since the value of $k_{\rm F}$ of Na is smaller than that of Al. 

To understand the phonon softening at the N point, we derive analytical expressions for the phonon frequency. From Eq.~(\ref{eq:dyn}), the frequencies at the N point are written as 
\begin{eqnarray}
\omega_1&=& \sqrt{\frac{4(A_1+A_2) + 2B_2}{M_{\rm ion}}},
\label{eq:omegaN1}
\\
\omega_2&=& \sqrt{\frac{4(A_1+A_2) + \frac{4}{3}B_1}{M_{\rm ion}}},
\label{eq:omegaN2}
\\
\omega_3 &=& \sqrt{\frac{4(A_1+A_2) + \frac{8}{3}B_1 + 2B_2}{M_{\rm ion}}},
\label{eq:omegaN3}
\end{eqnarray}
where $A_p$ and $B_p$ up to $p=2$ are considered because the use of two parameters $A_1(<0)$ and $B_1$ only is not enough to obtain a dynamically stable structure at $T_{\rm e}=0$ K. We emphasize that the expression for the lowest frequency $\omega_1$ in Eq.~(\ref{eq:omegaN1}) does not include the largest force constant $B_1$; The magnitude of $\omega_1$ is determined by a delicate balance between $A_1$, $A_2$, and $B_2$. Although $B_2$ increases with $T_{\rm e}$, the amount of the increase is completely cancelled out by the decrease in $A_1$ and $A_2$. The latter contributions are large enough to cause $\omega_{\rm 1}$ to decrease with $T_{\rm e}$. 

The polarization vectors $\bm{e}_1$, $\bm{e}_2$, and $\bm{e}_3$ corresponding to $\omega_1$, $\omega_2$, and $\omega_3$ in Eqs.~(\ref{eq:omegaN1})-(\ref{eq:omegaN3}), respectively, are written as
\begin{eqnarray}
\bm{e}_1=
\frac{1}{\sqrt{2}}
\left(
\begin{array}{c}
      0  \\
      1  \\
      -1 
    \end{array}
  \right), 
  \
  \bm{e}_2= 
\left(
\begin{array}{c}
      1  \\
      0  \\
      0 
    \end{array}
  \right), 
  \ 
    \bm{e}_3= 
    \frac{1}{\sqrt{2}}
\left(
\begin{array}{c}
      0  \\
      1  \\
      1 
    \end{array}
  \right).
  \nonumber\\
\end{eqnarray}
The vectors $\bm{e}_i$ ($i=1$ and $2$) and $\bm{e}_3$ are perpendicular to and parallel to the N point wavevector $\bm{q}=(0,1/2,1/2)$ in units of $2\pi/a_{\rm lat}$, respectively, which will be helpful to identify the soft mode experimentally. 

It should be noted that also for simple cubic structured lattices $B_1$ is not entered into the expression of the phonon frequencies at points X and M, implying an appearance of the phonon softening with $T_{\rm e}$. We thus speculate that the larger the NN coordination number $Z_{\rm C}$, the stronger the bond strength against the electron excitation. In fact, it has been shown that an electronic excitation can lead to a phonon softening in Bi with $Z_{\rm C}=6$ \cite{yan,murray} and Si with $Z_{\rm C}=4$ \cite{recoules,yan}, while it leads to a phonon hardening in hexagonal closed-packed structure of Mg with $Z_{\rm C}=12$ \cite{yan}.

%%%%%%%%%%%%%%%%%%%%%%%%%%%%%%%%%%%
\section{Summary}
We have studied the effect of the electron temperature on the phonon dispersion relations for fcc- and bcc-structured metals within a model pseudopotential approach. The phonon hardening and softening in simple metals are discussed in terms of the force constants and the adiabatic potential as a function of the electron temperature. The phonon hardening originates from a significant increase in the force constant for the first NN sites, while the phonon softening at the N point in bcc-structured metals originates from a delicate balance between force constants for the first and second NN sites.

The formulation of the present work can be extended to metals with $d$-electrons in a sense of the valence $Z$ change as discussed in the study of warm-dense gold \cite{fourment}, while parametrizing the relationship between $Z$ and $T_{\rm e}$ is necessary.

%Experimentally, the electron system can be excited by a femtosecond laser irradiation \cite{ernstorfer,siwick,jourdain}, yielding the nonequilibrium electron distribution. We assumed that the electron-electron collision rate is large enough to keep the quasiequilibrium distribution function with $T_{\rm e}$ \cite{allen}, while the validity of the quasiequilibrium approximation has been discussed recently \cite{muller,waldecker,ono2018}. The effect of the nonequilibrium distribution on the phonon properties enters through the response function defined in Eq.~(\ref{eq:response}) but the Fermi-Dirac distribution function in the integrand of Eq.~(\ref{eq:response}) should be replaced by the nonequilibrium distribution function, the time-evolution of which is governed by such as the Boltzmann equation coupled with the phonon distribution functions. Such a study enables us to predict how to observe the bond hardening in simple metals experimentally, while it is left for future works.

%For example, when $T_{\rm e}$ is increased up to 4.9 eV, the value of $Z$ for Au changes from 1.0 to 2.58 \cite{fourment}. This would lead to two competitive effects: One is an increase in the repulsive direct interaction $v_{\rm d}$ defined in Eq.~(\ref{eq:d}), and the other is an increase in the attractive electron-ion interaction $v_{\rm ps}$ defined in Eq.~(\ref{eq:Ashcroft}). In addition, the change in $Z$ causes a decrease in $r_s$ subject to the charge neutrality condition. 

\appendix
\section{Effective ion-ion interaction}
\label{app}
We outline the derivation of the ion-electron-ion interaction potential \cite{hartmann,grimvall} for the case of finite $T_{\rm e}$ by considering the effect of the electron-ion interaction to the total electron energy of the free-electron system. 

The Schr\"{o}dinger equation for the free-electron in a volume $\Omega$ is given by
\begin{eqnarray}
 H_0 \vert \bm{k} \rangle = \varepsilon (\bm{k}) \vert \bm{k} \rangle,
\end{eqnarray}
where $\varepsilon (\bm{k}) = \hbar^2k^2/(2m)$ is the free-electron energy and $\vert \bm{k} \rangle = e^{i\bm{k}\cdot \bm{r}}/\sqrt{\Omega}$ is the electron eigenstate. $\hbar$ is the Planck constant, $m$ is the bare electron mass $m$, and $\bm{k}$ is the wavevector. The Schr\"{o}dinger equation for the electron in a crystal with the number of the unit cell $N_{\rm c}$ is given as
\begin{eqnarray}
 (H_0 + W) \vert \Psi \rangle = E \vert \Psi \rangle,
\end{eqnarray}
where $W$ is a weak periodic potential, to which the electron-electron (e-e) interaction is also included, and written as 
\begin{eqnarray}
 W(\bm{r}) = \sum_{j} w(\bm{r} - \bm{R}_j)
\end{eqnarray}
with $w(\bm{r} - \bm{R}_j)$ being the potential energy of the lattice sites $\bm{R}_j$. Within the perturbation theory, the electron wavefunction is written as
\begin{eqnarray}
 \vert \Psi \rangle = 
 \vert \bm{k} \rangle  + \sum_{\bm{q}\ne 0} 
 \frac{\langle \bm{k} + \bm{q}\vert  W \vert \bm{k}\rangle}
 {\varepsilon (\bm{k}) - \varepsilon (\bm{k}+\bm{q})}
 \vert \bm{k} + \bm{q} \rangle,
 \label{eq:1st_wf}
 \end{eqnarray}
 while the electron eigenenergy is written as
 \begin{eqnarray}
 E(\bm{k}) &=& 
  \varepsilon (\bm{k}) + \langle \bm{k}\vert  W \vert \bm{k}\rangle
 \nonumber\\
 &+& \sum_{\bm{q}\ne 0} 
 \frac{\langle \bm{k} + \bm{q}\vert  W \vert \bm{k}\rangle 
 \langle \bm{k} \vert  W \vert \bm{k}+ \bm{q}\rangle}
 {\varepsilon (\bm{k}) - \varepsilon (\bm{k}+\bm{q})}.
 \label{eq:ek}
 \end{eqnarray}
The total energy per a unit cell is then given as
\begin{eqnarray}
E_{\rm tot}
&=&  \frac{2}{N_{\rm c}} \sum_{\bm{k}} f(\bm{k};T_{\rm e}) E(\bm{k})
 + E_{\rm dc},
 \label{eq:total_energy}
\end{eqnarray}
where $f(\bm{k};T_{\rm e})$ is the Fermi-Dirac distribution function for the energy $E(\bm{k})$ and the temperature $T_{\rm e}$, which is introduced in this work to account for the electron excitation. The prefactor 2 in the first term in Eq.~(\ref{eq:total_energy}) comes from spin degeneracy. The second term $E_{\rm dc}$ accounts for the double-counting of the e-e interaction energy. As usual, we decompose $E_{\rm tot}$ into 
\begin{eqnarray}
E_{\rm tot}=E_{\rm free}+ E_{\rm bs},
\end{eqnarray}
where $E_{\rm free}$ is the free-electron energy that is independent of the ion position, while $E_{\rm bs}$ is the bandstructure energy that depends on the geometrical configuration of the ions. $E_{\rm bs}$ comes from both the last term of Eq.~(\ref{eq:ek}) and $E_{\rm dc}$, and can be expressed as 
\begin{eqnarray}
 E_{\rm bs} = \sum_{\bm{q}\ne 0} S^{*}(\bm{q}) S(\bm{q}) F(\bm{q})
 \end{eqnarray}
with the structure factor
\begin{eqnarray}
 S(\bm{q}) = \frac{1}{N_{\rm c}} \sum_{j} e^{-i\bm{k}\cdot \bm{R}_j}
 \label{eq:structure_factor}
\end{eqnarray}
and the energy-wavenumber characteristic $F(\bm{q})$. By using Eq.~(\ref{eq:structure_factor}), one obtains
 \begin{eqnarray}
 E_{\rm bs} 
 &=& 
 \frac{1}{N_{\rm c}^{2}}\sum_{\bm{q}\ne 0} \sum_{i}\sum_{j}
 e^{i\bm{q}\cdot (\bm{R}_i-\bm{R}_j)} F(\bm{q})
 \nonumber\\
 &=& 
 \frac{1}{2N_{\rm c}}\sum_{i \ne j}
 v_{\rm ind}(\vert \bm{R}_i-\bm{R}_j \vert)
 + \frac{1}{N_{\rm c}} \sum_{\bm{q}\ne 0} F(\bm{q}),
 \nonumber\\
 \end{eqnarray} 
where $v_{\rm ind}(\vert \bm{R}_i-\bm{R}_j \vert)$ is the indirect interaction potential between ions at $\bm{R}_i$ and $\bm{R}_j$ and expressed as
  \begin{eqnarray}
  v_{\rm ind}(\vert \bm{R} \vert)
&=&
\frac{2}{N_{\rm c}} \sum_{\bm{q}\ne 0} 
 e^{i\bm{q}\cdot (\bm{R}_i-\bm{R}_j)} F(\bm{q})
 \nonumber\\
 &=&
 \frac{\Omega_{\rm a}}{\pi^2} 
 \int_{0}^{\infty} dq q^2 \frac{\sin (qR)}{qR}F(q)
 \label{eq:ind_app}
 \end{eqnarray} 
with $\Omega_{\rm a} = \Omega/N_{\rm c}$. We assumed that $F$ depends on the magnitude of $\bm{q}$ only. When the contributions from the Hartree, exchange, and correlation interactions are included to $W$, $F(q)$ can be written as \cite{grimvall}
\begin{eqnarray}
F(q) =
- \frac{\epsilon_0 q^2 }{2e^2\Omega_{\rm a}}  
  \frac{\chi (q, T_{\rm e}) v_{\rm ps}^2(q)}
  {1+\left[1-G(q)\right] \chi(q, T_{\rm e})}. 
  \label{eq:fq}
 \end{eqnarray}
By substituting Eq.~(\ref{eq:fq}) into Eq.~(\ref{eq:ind_app}), we obtain the expressions in Eqs.~(\ref{eq:ind}) and (\ref{eq:ind2}). 

%{\color{red} This assumption is validated within the random-phase-approximation discussed below.}
 
%%%%%%%%%%%%%%%%%%%%%%%%%%%%%%%%%%%


\begin{thebibliography}{99}

\bibitem{lin} Z. Lin, L. V. Zhigilei, and V. Celli, Electron-phonon coupling and electron heat capacity of metals under conditions of strong electron-phonon nonequilibrium, Phys. Rev. B {\bf 77}, 075133 (2008).

\bibitem{recoules} V. Recoules, J. Cl\'{e}rouin, G. Z\'{e}rah, P. M. Anglade, and S. Mazevet, Effect of Intense Laser Irradiation on the Lattice Stability of Semiconductors and Metals, Phys. Rev. Lett. {\bf 96}, 055503 (2006).

\bibitem{ernstorfer} R. Ernstorfer, M. Harb, C. T. Hebeisen, G. Sciaini, T. Dartigalongue, R. J. D. Miller, The Formation of Warm Dense Matter: Experimental Evidence for Electronic Bond Hardening in Gold, Science {\bf 323}, 1033 (2009).

\bibitem{leguay} P. M. Leguay, A. L\'{e}vy, B. Chimier, F. Deneuville, D. Descamps, C. Fourment, C. Goyon, S. Hulin, S. Petit, O. Peyrusse, J. J. Santos, P. Combis, B. Holst, V. Recoules, P. Renaudin, L. Videau, and F. Dorchies, Ultrafast Short-Range Disordering of Femtosecond-Laser-Heated Warm Dense Aluminum, Phys. Rev. Lett. {\bf 111}, 245004 (2013).

\bibitem{cho} B. I. Cho, T. Ogitsu, K. Engelhorn, A. A. Correa, Y. Ping, J. W. Lee, L. J. Bae, D. Prendergast, R. W. Falcone, and P. A. Heimann, Measurement of Electron-Ion Relaxation in Warm Dense Copper, Sci. Rep. {\bf 6}, 18843 (2016).

%\bibitem{jourdain} N. Jourdain, L. Lecherbourg, V. Recoules, P. Renaudin, and F. Dorchies, Electron-ion thermal equilibration dynamics in femtosecond heated warm dense copper, Phys. Rev. B {\bf 97}, 075148 (2018).

\bibitem{dorchies} F. Dorchies, V. Recoules, J. Bouchet, C. Fourment, P. M. Leguay, B. I. Cho, K. Engelhorn, M. Nakatsutsumi, C. Ozkan, T. Tschentscher, M. Harmand, S. Toleikis, M. St\"{o}rmer, E. Galtier, H. J. Lee, B. Nagler, P. A. Heimann, and J. Gaudin, Time evolution of electron structure in femtosecond heated warm dense molybdenum, Phys. Rev. B {\bf 92}, 144201 (2015).

\bibitem{groth} S. Groth, T. Dornheim, T. Sjostrom, F. D. Malone, W. M. C. Foulkes, and M. Bonitz, Ab initio Exchange-Correlation Free Energy of the Uniform Electron Gas at Warm Dense Matter Conditions, Phys. Rev. Lett. {\bf 119}, 135001 (2017).

\bibitem{minakov} D. V. Minakov and P. R. Levashov, Melting curves of metals with excited electrons in the quasiharmonic approximation, Phys. Rev. B {\bf 92}, 224102 (2015).

\bibitem{yan} G. Q. Yan, X. L. Cheng, H. Zhang, Z. Y. Zhu, and D. H. Ren, Different effects of electronic excitation on metals and semiconductors, Phys. Rev. B {\bf 93}, 214302 (2016).

\bibitem{harbour} L. Harbour, M. W. C. Dharma-wardana, D. D. Klug, and L. J. Lewis, Equation of state, phonons, and lattice stability of ultrafast warm dense matter, Phys. Rev. E {\bf 95}, 043201 (2017).
 
 \bibitem{bottin} F. Bottin and G. Z\'{e}rah, Formation enthalpies of monovacancies in aluminum and gold under the condition of intense laser irradiation, Phys. Rev. B {\bf 75}, 174114 (2007).
 
\bibitem{hartmann} W. M. Hartmann and T. O. Milbrodt, Model-Potential Calculations of Phonon Energies in Aluminum, Phys. Rev. B {\bf 3}, 4133 (1971).

\bibitem{mermin} N.W. Ashcroft, N. D.Mermin, and D.Wei, {\it Solid State Physics}, revised edition, (Cengage, Boston, 2016).

\bibitem{ashcroft} N. W. Ashcroft, Phys. Lett. {\bf 23}, 48 (1966).

\bibitem{UI} S. Ichimaru and K. Utsumi, Analytic expression for the dielectric screening function of strongly coupled electron liquids at metallic and lower densities, Phys. Rev. B {\bf 24}, 7385 (1981).

%\bibitem{siwick} B. J. Siwick, J. R. Dwyer, R. E. Jordan, and R. J. D. Miller, An atomic-level view of melting using femtosecond electron diffraction, Science {\bf 302}, 1382 (2003).

\bibitem{murray} \'{E}. D. Murray, S. Fahy, D. Prendergast, T. Ogitsu, D. M. Fritz, and D. A. Reis, Phonon dispersion relations and softening in photoexcited bismuth from first principles, Phys. Rev. B {\bf 75}, 184301 (2007).

\bibitem{fourment} C. Fourment, F. Deneuville, D. Descamps, F. Dorchies, S. Petit, O. Peyrusse, B. Holst, and V. Recoules, Experimental determination of temperature-dependent electron-electron collision frequency in isochorically heated warm dense gold, Phys. Rev. B {\bf 89}, 161110(R) (2014).

%\bibitem{allen} P. B. Allen, Theory of Thermal Relaxation of Electrons in Metals, Phys. Rev. Lett. {\bf 59}, 1460 (1987).

%\bibitem{muller} B. Y. Mueller and B. Rethfeld, Relaxation dynamics in laserexcited metals under nonequilibrium conditions, Phys. Rev. B {\bf 87}, 035139 (2013).

%\bibitem{waldecker} L. Waldecker, R. Bertoni, and R. Ernstorfer, and J. Vorberger, Electron-Phonon Coupling and Energy Flow in a Simple Metal beyond the Two-Temperature Approximation, Phys. Rev. X {\bf 6}, 021003 (2016).

%\bibitem{ono2018} S. Ono, Thermalization in simple metals: Role of electron-phonon and phonon-phonon scattering, Phys. Rev. B {\bf 97}, 054310 (2018).

\bibitem{grimvall} G. Grimvall, {\it The Electron-Phonon Interaction in Metals}, (North-Holland, Amsterdam, 1981).

\end{thebibliography}
\end{document}